\documentclass[12pt]{article}
\usepackage{amsmath}
\usepackage[psamsfonts]{amssymb}
\usepackage{graphicx}

\begin{document}

\begin{flushleft} \footnotesize{arXiv.org e-print archive,
hep-th/0205196 (http://arXiv.org/abs/hep-th/0205196) }
\end{flushleft}

\begin{center}
\section*{\bf Gravitation as deduced from
submicroscopic quantum mechanics}
\end{center}

 \vspace{4mm}

\begin{center}
{\bf Volodymyr Krasnoholovets}
\end{center}

\begin{center}
{Institute of Physics, National Academy of Sciences, \\ Prospect
Nauky 46,   UA-03028 Ky\"{\i}v, Ukraine } \\ (web page
http://www.inerton.kiev.ua)

\vspace{4mm} \hspace{15 cm}16 May 2002, 4 April 2003, 23 April
2004, 25 April 2005

\end{center}

\vspace {2mm} {\small

\begin{center}
{\bf Abstract}
\end{center}

Based on the model of a "soft" cellular space and deterministic
quantum mechanics developed previously, the scattering of a free
moving particle by structural units of the space -- superparticles
-- is studied herein. The process of energy and inert mass
transmission from the moving particle to superparticles and hence
the creation of elementary excitations of the space -- inertons --
are analyzed in detail. The space crystallite made up around a
particle in the degenerate space is shown to play the key role in
those processes. A comprehensive analysis of the nature of the
origin of gravitation, the particle's gravitational potential
$1/r$, and the gravitational interaction between material objects
is performed. It seems reasonably to say that the main idea of the
work may briefly be stated in the words: No motion, no gravity.\\

{\bf  Key words}: space, gravitation, inertons, mass dynamics,
quantum theory     \\

 {\bf PACS:} 03.65.Bz Foundations, theory of
measurement, miscellaneous theories; \\    03.75.-b Matter waves;
\  04.60.-m Quantum gravity \vspace{6mm} }

\begin{flushright} \footnotesize{The development of knowledge brings
other results  \kern 1pt \  \ \        \\    \ \ \ \  than those
ignorance cultivates. \ This we learned  \ \ \ \ \          \\
from sages. \kern 1pt  \ \ \ \ \ \ \ \ \ \ \ \ \ \ \ \ \ \ \ \ \ \
\ \ \ \ \ \ \ \ \ \ \ \ \ \ \ \ \ \ \ \ \ \ \ \ \ \ \ \ \ \ \

\hspace{5.3cm} \ \break    $\acute{S}r\bar{\imath}$
$\bar{I}\acute{s}opanishad$, \ Mantra ten \ \ }
\end{flushright}

\section{\bf  Introduction}

 \hspace*{\parindent}
Gravitation still remains the most obscure theme in physics.
Notwithstanding this, the conceptual foundations of general
relativity and quantum mechanics allow concrete modifications in
each other in the interface region, Ahluwalia [1]. Describing
conceivable new generation quantum-gravity experiments in certain
atomic systems, Ahluwalia [2] then notes that they are based on
the possibility that quantum gravity might affect the nature of
general symmetry, or that the theory of general relativity itself
may not provide a complete description of gravitation.

   Many fundamental problems of gravity and the quantum behavior
of the matter have been raised also in remarkable review by Sorkin
[3]. In review [4], Dowker and Sorkin investigate some quantum
properties of spatial topological geons, particles in 3+1 quantum
gravity. Geons allow the construction of spin statistics both
fermionic and bosonic and this seems to be a way to an
understanding the inner construction of Nature. It is interesting
that the authors note that geons are found in some common
environment. But what kind of environment? May geons interact with
the environment? In particular, what is the law of the motion of
the geon? Of course, these and other questions need separate
studies. However, the fact of the presence of an environment has
engaged a special attention. This signifies that the spacetime, or
a vacuum is endowed with structural properties that turn us out to
a certain aether, but rather quantum one.

  In recent years several serious attempts to re-introduce an aether
in physics have been made [5-11]. Similarly, in the previous works
of the author [12-19] a simulation of a vacuum in the form of a
cellular elastic space has been proposed and deterministic quantum
mechanics based on the strong interaction of a moving particle
with such a space net has been constructed. The real space has
been suggested to consist of peculiar super densely packed
superparticles, or balls, which are found in the degenerate state
over all the multiplets. A particle is created from a
superparticle and hence the particle is interpreted as a local
deformation of the degenerate space. Such idea agrees very well
with the mathematical study of space carried out by Bounias and
Bonaly [20,21]  (see also Ref. [22]) who have shown basing on
topology and set theory that the necessity of the existence of the
empty set leads to the topological spaces resulting in a "physical
universe". This allows the investigation of links between physical
existence, observability and information. Thus the empty hyperset
provides for a formal structure that correlates with the
degenerate cell of space and supports conditions for the existence
of a universe. Magnon [23] following Bounias also pointed out
about some "primordial cell" and "existential principle".

Moreover, the surprising thing is that the information on the
existence of the supreme substance and the first cause of matter
in the form of an indivisible thing is contained in
$Bhagavad$-$g{\bar \imath}t{\bar a}$ (see also Bhaktivedanta Swami
Prabhupada [24]): "In spite of that material body is subjected to
destruction, [the subtle particle] is eternal" [{\it Bg.} 2.18];
"It never takes birth and never dies at any time nor does it come
into being again when the material body is created. It is
birthless, eternal, imperishable and timeless and is inviolable
when the body is destroyed" [{\it Bg.} 2.20]; "After some time it
is disenthralled by entirely annihilation of the material body.
Yet it endures the destruction of the material world" [{\it Bg.}
2.22]; "It is not fissionable, not burning out, not soluble, and
not drying up" [{\it Bg.} 2.23]; "Since it is not visible, its
entity does not change, its properties remain unchangeable" [{\it
Bg.} 2.24]; "Yet there is another nature, which is eternal and is
transcendental to this manifested and unmanifested matter. It is
supreme and is never annihilated. When all in this world is
annihilated, that part remains as it is" [{\it Bg.} 8.20].

There are also other approaches introducing an aether and those
ones that aims to the examination of properties of space-time at
very short distances. In particular, Amelino-Camelia notes that
the nature of space-time has to show a "fuzziness" at distances
close to the Planck one [25] and at these distances particles may
be described as geometry "defects" [26]. "Foamy" quantum gravity
fluctuations are studied in Refs. [27,28] and such fluctuations
seems can be able to modify particle propagation in an observable
way [29]. Semi-classical space-time is emerged in canonical
quantum gravity in the loop presentation as a polymer-like
structure at microscales, which allows a possible correction to
the Maxwell equations caused by quantum gravity [30]. Matone [31]
has studied the quantum Hamilton-Jacobi equation for a system of
two particles and concluded that gravitation in such system had a
pure quantum-mechanical origin.

Coming back to the author's concept of the constitution of space
and the creation of elementary particles in it [12-19], we should
note that the approach supposes the formation of a deformation
coat around the created particle [12-16]. The coat differs from
the degenerate space in that that its superparticles possess mass.
Therefore the coat may be called the space crystallite. Its size
corresponds to the Compton wavelength $\tilde \lambda_{{\kern
1pt}0}$ of the created particle, $\tilde \lambda_{{\kern
1pt}0}=h/M_0c$ [14]. According to the definition [12,13], the
induction of mass means that the volume of a superparticle changes
from its volume in the degenerate space. In other words, if we set
that a superparticle constricts with deformation, the mass will be
defined as the ratio of superparticle's initial and final volumes,
\ $m_0\propto {\cal V}/V_0^{{\kern 0.5pt}\rm sup}$ \ for the
massive superparticle and \ $M_0 \propto {\cal V}/V_0^{{\kern
0.5pt}\rm par}$ \ for the particle, where ${\cal V}$ is the
typical volume of a degenerate superparticle and the volumes of a
deformed superparticle and a particle are respectively
$V_0^{{\kern 0.5pt}\rm sup}$ and $V_0^{{\kern 0.5pt}\rm par}$.
When a particle moves, the crystallite travels together with it.
However superparticles themselves are motionless: the crystallite
state migrates by a relay mechanism. The rearrangement of
superparticles due to the particle's motion takes place with a
velocity that equals or exceeds the speed of light $c$, but the
motion of the particle itself occurs with the velocity $v_0<c$
(hereinafter $v_0$ designates the initial particle's velocity,
which the particle acquires at a momentary push). The moving
particle emits and absorbs elementary excitations of the space --
inertons, which appear as a result of friction that the particle
undergoes when moves against superclosely packed superparticles.
Why do inertons not leave the particle totally? Why do they come
backwards again? This is because they carry not only mass but
electromagnetic polarization as well as particles are charged.
However, this is the subject of a separate study. Here we only
point out that the velocity $c_{\rm free}$ of an absolutely free
inerton that might migrate in the degenerate space should exceed
the velocity of light $c$, perhaps several times. Accordingly, the
initial velocity $\hat c$ of electromagnetic polarized inertons at
which they are emitted from the moving particle and accompany it,
is not able to reach the threshold value $c_{\rm free}$ and
probably $c < \hat c < c_{\rm free}$. At the same time the mean
value of the velocity of the particle's inertons over a period
still remains less then the velocity of light $c$ even when the
velocity $v_0$ of the particle approaches to $c$ (in fact $\frac
12 (\hat c + 0) < c$ even at $\hat c = \sqrt{c^2 + c^2}$, see
expression (7) below).

Detailed theoretical consideration of the motion of a canonical
particle has shown [12-14] that owing to the interaction with
superparticles the particle looses its kinetic energy on the
section $\lambda/2$ of the particle path where $\lambda$ is the
amplitude of spatial oscillations of the particle (the de Broglie
wavelength).  The lost energy is spent on the creation of
inertons.  On the next section $\lambda/2$ the particle absorbing
inertons acquires the velocity $v_0$, and so on. Thus inertons
form a substructure of the matter waves. It should be emphasized
that the major theoretical prediction, the existence of clouds of
inertons surrounding particles such as electrons and atoms,
indeed, has recently been substantiated in a number of experiments
[17-19].

   In addition to friction, or inertia of the space, which results
into the generation of inertons, the moving particle undergoes the
dynamic pressure on the side of the whole space [13] (the pressure
acts on the particle through its coat). The space pressure causes
an additional deformation in the particle volume, $V_0^{\rm
part}\rightarrow V_0^{\rm part}\sqrt{1-v^{\kern 1pt 2}_0/c^{\kern
1pt 2}}$, \ that is, the pressure induces the  so-called
relativistic mass $M=M_0/\sqrt{1- v^2_0/c^{\kern 1pt 2}}$. Here we
write the speed of light $c$, because the electromagnetic
polarization that accompanies any particle imposes a limitation on
the speed of inertons. Since the value of $v_0$ varies in the
spatial interval $\lambda$ of the particle path, and this is one
of the peculiarities of the model constructed, the particle energy
should periodically change as well. The energy passing from the
particle to its inerton cloud is given by the kinetic energy of
the particle $\frac 12 Mv^2$. At this moment the particle mass
changes from $Mc^2$ to $Mc^2-\frac 12 Mv^2$ and hence the behavior
of the mass obeys the law
  $$
M \to M_0 \to M \to M_0 \to ...
  $$
That is, the value of mass oscillates along the particle path within the spatial period, or amplitude $\lambda$.

De Broglie [32] was the first to indicate that the corpuscle
dynamics was the basis for the wave mechanics. With the
variational principle, he obtained and studied the equations of
motion of a massive point reasoning from the typical Lagrangian
  $$
{\cal L}= -M_0{\kern 1pt}c^{\kern 1pt
2}\surd\overline{1-v^2/c^{\kern 1pt 2}}
  $$
in which, however, the velocity $v$ of the point was constant
along a path. The study showed that the dynamics had the
characteristics of the dynamics of the particles with a variable
proper mass. Oudet [33] conjectured similar peculiarities for the
electron. Papini and Wood [34] studied a geometrical solution to
the de Broglie variable mass problem. What is more, the de Broglie
view is well substantiated as it immediately follows from the
Schr\"odinger equation: the equation giving quantum solutions
contains a pure classical parameter -- the unchangeable particle
mass.

    Based on the theory developed in recent author's works [12-19],
the present paper shows that the creation of inertons from a
moving particle is stipulated by the existence of the space
crystallite around the particle. The de Broglie's idea and the
author's previous results are taken as the starting point. The
study includes an extensive description of the processes of the
energy and mass transmission from the particle to superparticles
when the particle and coming superparticles collide. Besides for
the first time the theory arises the question, how is the
gravitational potential induced by a particle/object in the
ambient space? It is read that such induction is caused by
inertons enclosing any material object. Thereby it is the dynamic
inerton field that is responsible for the generation of the static
Newton potential $1/r$. The appearance of the gravitational
interaction between both particles and material objects is
elucidated in some detail.

\section {\bf  Emission of inertons }

\hspace*{\parindent} For the solution of equations of motion of
the particle and inertons the relation
\begin{equation}
M{\dot X}^2_i=m_{{\kern 0.5pt}i}{\hat c}^{\kern 1pt 2} \label{1}
\end{equation}
has been used in the preceding papers of the author [12-14]. In
(1) $M$ is the particle mass, ${\dot X}_i$ is the vector of the
particle velocity at the moment of the $i$th inerton emission
($X_i$ is the radius vector of the particle and the dot over $X_i$
means the differentiation in respect to the proper time of the
particle), $m_{{\kern 0.5pt}i}$ is the mass of the $i$th inerton
and $\hat c$ is its initial velocity. Relation (1) is the
consequence of the intersection of geodesics of the particle and
the $i$th inerton. At the moment of the $i$th collision of the
particle and the superparticle, the former emits the inerton, as
follows from (1), whose energy is equal  to the double kinetic
energy of the particle itself. It turns out that a moving object
emits an enormous quantity of inertons $N$ within a half-period of
its spatial oscillation ($N=\lambda / {\cal V}^{1/3}$ where
$\lambda$ is the spatial period identical to the de Broglie
wavelength and ${\cal V}^{1/3} \sim 10^{-30}$ m, or ${\cal
V}^{1/3} \sim l_{\rm Planck} \approx 10^{-35}$ m,  is the
suggested size of the superparticle). So the energy of the emitted
inerton cloud is of the order of the huge magnitude $NMv_0^2/2$.

   This situation is possible when the $i$th inerton gains the energy
because the medium around the particle is in the vibrating state
rather than because of loss in kinetic energy of the particle
itself (its energy suffices for the creation of one inerton only
[12]). The availability of the crystallite in the space around the
particle has been proven theoretically in paper [14]. The
crystallite's superparticles are characterized by mass. Hence
collective vibrations of massive superparticles, similarly to
vibrations of atoms in an ordinary solid crystal, should be
inherent to the crystallite. Therefore, vibrating superparticles
are able to strike the particle in such a way that the particle
will loss energy in collisions and then will generate excitations
in the surrounding. Excitations caught by the vibrating
superparticles will carry away from the particle.

At the formal consideration of quantum gravity, for instance in
the model of D-brane string solutions, researchers also face the
problem of scattering. Kabat and Pouliot [35] treated the
zero-brane dynamics in which one could probe distances much
shorter than the string space. Ellis et al. [36] have studied the
gravitational recoil effects induced by energetic particles. In
their approach defects in space-time are derived as an
energy-dependent refractive index and D-brane foam has
corresponded to a minimum-uncertainty wave-packet. The shift of
D-brane particle has been induced by the scattering after
interaction with a closed-string state.

Let us treat now the process of the particle scattering by
superparticles in the model discussed in detail.

Two relationships were previously derived [13,14]:
\begin{equation}
\Lambda = \lambda {\kern 1pt}{\hat c}/v_0, \label{2}
\end{equation}
\begin{equation}
\Lambda = {\tilde \lambda}_{{\kern 1pt}v_{_0}}{\hat{c}}^{\kern 1pt
2}/v_0^2 . \label{3}
\end{equation}
Here expression (2) connects the spatial period $\lambda$ of the
particle oscillations (the de Broglie wavelength) with the
specific enveloping amplitude $\Lambda$ of the inerton ensemble.
Expression (3) connects the effective size of the crystallite
${\tilde \lambda}_{{\kern 1pt}v_{_0}}$, defined by the Compton
wavelength of the particle, with the specific amplitude $\Lambda$
of inertons.

    However the crystallite is dynamic: along the particle's velocity
vector permanently occurs the relay readjustment of superparticles
from the massless to massive state and again to the massless one
(Fig. 1).
\begin{figure}
\begin{center}
\includegraphics[scale=2.2]{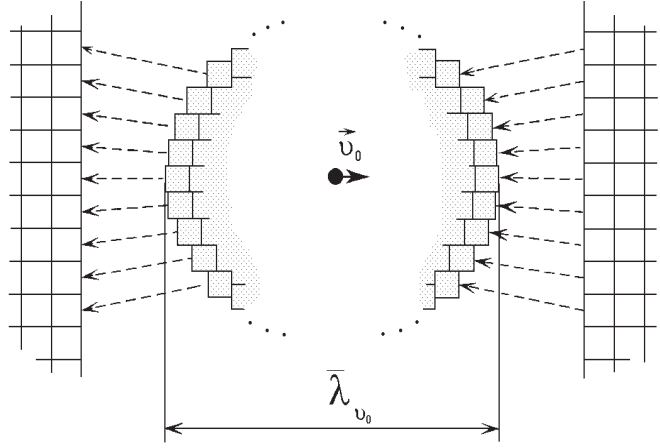}
\caption{Rearrangement of the mass state of superparticles around
the moving particle.} \label{Figure 1}
\end{center}
\end{figure}
 The resultant of the movement of superparticles is
directed antiparallel to the vector $\vec v_0$. Thus along the
line of the particle motion the state of superparticles changes
dynamically: all the time in the coarse of the particle motion
superparticles change their state from the degenerate one to the
massive one. But in transversal directions the state of
superparticles remains practically unaltered: superparticles
surrounding the particle continuously save the same massive state,
i.e., in these directions superparticles in the crystallite might
be considered as hard. Since in any crystal atoms vibrate, massive
superparticles should vibrate in the said crystallite as well.
However from the pattern above it turns out that superparticles in
the crystallite suffer rather pure transversal vibrations -- their
equilibrium positions transversely vibrate in reference to the
vector $\vec v_0$. Such transversal vibrations of the crystallite
we may call the transversal vibrating mode.

The elasticity constant $\gamma$ of the crystallite and the
average mass $m_{\rm cr}$ of a crystallite's superparticle
determine the cyclic frequency of these collective vibrations [14]
\begin{equation}
\omega_{k_{v_0}}=\sqrt{\gamma/m_{{\kern 0.5pt}\rm cr}}. \label{4}
\end{equation}
Note that the value of $\gamma$ is given by the particle mass at
rest $M_0$, so $\gamma$ is not universal.

   Now one can reproduce in detail the picture of the particle motion
with the emission and absorption of inertons. Let us assume that
at some initial moment, the particle has the velocity $v_0$ and
mass $M=M_0/\sqrt{1-v_0^{{\kern 0.5pt}2}/c^{{\kern 0.5pt}2}}$,
equivalently, the particle is characterized by the additional
deformation to its own volume $V_0^{\rm part}$ due to the received
velocity $v_0$: \ $V_0 \rightarrow V^{\rm part} = V_0^{\rm
part}\sqrt{1-v^{{\kern 0.5pt}2}_0/c^{{\kern 0.5pt}2}}$. The
particle in motion runs into superparticles of the crystallite,
which vibrate in directions transversal to the vector $\vec v_0$.
The superparticles' vibratory longwave modes can be considered as
a longwave excitation of the crystallite:
\begin{equation}
        k_{\kern 1pt {v_0}} a\ll 1
\label{5}
\end{equation}
where $k_{\kern 1pt {v_0}}=\pi/{\tilde \lambda}_{{\kern
1pt}v_{_0}}$ is the wave number and $a$ is the size of a
superparticle in the crystallite. The longwave approximation (5)
is true not only for $v_0 \ll c$ but also for the velocity $v_0$
close to $c$. The violation of inequality (5) happens only at the
up-relativistic velocity $v_0$ of the particle when ${\tilde
\lambda}_{{\kern 1pt}v_{_0}}$ is brought near to $a$ (but this
value $v_0$ is certainly unattainable experimentally). Note that
here we have the absolute analogy with the solid because it is
known from the theory of solid state (see, e.g. Ref. [37]) that
longwave excitations are elastic waves of the medium. The velocity
of these waves  -- the sound velocity $v_{\rm sound}$ -- is
determined from the relation $v_{\rm sound} = \bar a \sqrt{\bar
\gamma/\bar M}$, where $\bar a$ is the lattice constant, $\bar M$
is the atom mass and $\bar \gamma$ is the elastic constant of a
crystal.  In this approximation, the group and phase velocities
coincide.

   The vibration energy saved in the crystallite of a moving particle
can be found from the equation [14]
\begin{equation}
\hbar{\kern 2pt}\omega_{k_{v_{_0}}}=Mc^{\kern 1pt 2} \label{6}
\end{equation}
where $M=M_0/\sqrt{1-v_0^2/c^{\kern 1pt 2}}$ is the total mass (we
ignore the energy that stemming from the particle spin, see Ref.
[14] for details). Essentially, collisions of the particle with
the crystallite mode represent an impetuous attack (with the speed
of $c$) to which the particle is subjected on the part of the
front of this mode. Each of the $i$th action of collisions results
in the emission of the $i$th inerton: the mode knocks out the
inerton from the particle. The $i$th inerton has a mass $m_i$ and
two velocity components: ${\dot x}^{\perp}_i$ directed
perpendicular to the vector ${\vec v}_0$ and ${\dot
x}^{\parallel}_i$ along this vector. It is obvious that the
component ${\dot x}^{\parallel}_i$ passes over from the particle
to the $i$th inerton; the component ${\dot x}^{\perp}_i$ is caused
by the momentum of the crystallite mode. Supposing the speed  $c$
is the velocity of the crystallite mode, we get relation
\begin{equation}
\sqrt{({\dot x}^{\parallel}_i)^2 + ({\dot x}_i^{\perp})^2}=\hat c, \quad {\rm or} 
 \quad  \sqrt{({\dot x}^{\parallel}_i)^2 + c^{\kern 1pt 2}}=\hat c.
\label{7}
\end{equation}
Let $m_{{\kern 0.5pt}i}{\hat c}^{\kern 1pt 2}$ be the total energy
of the $i$th inerton. After a series of $N$ collisions of the
particle with the mode, the energy of the latter should gradually
decrease: $\hbar{\kern 1.5pt}\omega_{{v_{0}}}
>\cdots >\hbar{\kern 1.5pt}\omega_{{\kern 0.5pt}i}>\hbar{\kern 1.5pt}
\omega_{{\kern 0.5pt}i+1}>\cdots >\hbar {\kern 1.5pt} \omega_N$.
Therefore, the mode spends the energy $\hbar{\kern
1pt}\omega_{{\kern 0.5pt}i}$ for splitting of the particle and the
creation of the inerton that appears with the same energy
$\hbar{\kern 1.5pt}\omega_{{\kern 0.3pt}i} = m_{{\kern
0.5pt}i}{\kern 1pt}{\hat c}^{\kern 1.5pt 2}$. Immediately after
the $i$th collision the whole space net readjusts the crystallite
to a new quasi-equilibrium state in which the vibration energy
$\hbar{\kern 1pt}\omega_{{\kern 0.5pt} i+1}$ corresponds to a new
total particle energy $M_{i+1}{\kern 1pt}{\hat c}^{{\kern 1pt}2}$.
And then again a collision occurs and the energy passes
from the crystallite mode to the $(i+1)$th inerton, etc. until the
particle stops after the $N$th collision, i.e., when its velocity
and inert mass are exhausted, ${\dot X} \rightarrow 0$, $M_i
\rightarrow M_0$.

However, how is the inerton mass created? Obviously the process of
mass creation takes place at the sacrifice of the particle mass.
It should be assumed that at the $i$th collision of the particle
with the mode, the particle loses a portion $\delta V_i$ of its
initial relativistic deformation $V_0^{\rm
part}\sqrt{1-v_0^2/c^{\kern 1pt 2}}$, or, in other words, the
relativistic mass $M$ decreases on a value of the inerton mass
$m_i$  created on the $i$th coming superparticle. Thus the
particle mass also acquires the index $i$ and the generalization
for expression (1) should be the relationship
\begin{equation}
M_i{\dot X}^2_i =m_i{\hat c}^{\kern 1pt 2}. \label{8}
\end{equation}

     Now let us dwell on the inerton motion. The inerton is created
in the elastic medium, i.e. crystallite, which is specified by the
elasticity constant $\gamma$. The constant $\gamma$ determines the
interaction of the created inerton with the elastic force of the
crystallite. As a result of such an interaction, the elastic force
aspires to make the inerton return to the particle. Thereby the
force sets the inerton into oscillation along the line which is
superimposed with the vector of the initial inerton velocity $\vec
{\hat c}$. The cycle frequency of oscillation of the $i$th inerton
is
\begin{equation}
\omega_i=\sqrt{\gamma/m_i}.
\label{9}
\end{equation}
The maximum distance to which the inerton migrates from the
particle is the amplitude $\Lambda_i$, or $\Lambda_i={\hat
c}/(\omega_i/2\pi)$. Behind the crystallite boundary (in the case
when $v_{{\kern 0.5pt}0}<c$, i.e., when $\Lambda_i$ prevails the
Compton wavelength $\tilde\lambda_{{\kern 1pt}v_{_0}}$) the
inerton is guided by the degenerate space whose elasticity is
adjusted to the mass of the moving inerton in conformity with
relation (9), i.e., $\gamma_{{\kern 0.5pt}\rm space}=\gamma
=\omega^{\kern 1pt 2}_{{\kern 0.5pt}i}{\kern 1pt} m_{{\kern
0.5pt}i}$. Note once again that $\gamma$ is not a universal
parameter of the universe. This is a consequence of the axiom
[12,13] of adiabatic motion of particles/quasi-particles in space
when a moving object does not leave faults into the passed
range of the cellular space. In this event spatial oscillations of
the particle/quasi-particle are exemplified by the adiabatic
invariant of space, Planck's constant $\hbar$, and in the case of
our inerton this adiabatic invariant can be written as
$\hbar=(\frac 12 {\kern 1pt} m_i{\kern 1pt}{\dot x}_i^{\kern 1pt
2})/\omega_i$.

On the other hand, in papers [12-14] we characterized the $i$th
inerton by the frequency of its collisions with the particle,
$1/T_i$. Therefore we can relate this frequency to the oscillation
frequency (9), i.e. $\omega_i=2\pi/2T_i$. The parameter $1/T_i$
provided for the energy exchange between the particle and the
$i$th inerton. It is this parameter that allowed us to obtain the
periodicity in solutions for dynamic variables of the particle and
inertons. According to these solutions, within even time
half-period of collisions $T_i/2$ and the spatial one
$\lambda_{{\kern 1pt}i}/2$, the particle is scattered by the
crystallite vibrations with the subsequent formation of inertons.
Within odd half-period the inertons are absorbed by the particle;
they transmit the mass and the longitudinal component of the
velocity to the particle, i.e., the inertons guide the particle.
The guided particle is followed by the restoration of the
vibrating crystallite mode, or in other words, the whole
degenerate space restores the crystallite state to the initial
dynamic state at which the particle possesses the initial velocity
vector $\vec v_0$.

    We have stressed that the energy of the emitted inerton is
in the direct proportion to the particle energy at the moment of
collision with the crystallite mode. Since both the energy of the
particle and that of the mode decrease from collision to
collision, the same relation  should be true for the energy of
inertons. It follows that the inequality $m_{{\kern 0.5pt}i+1} <
m_{{\kern 0.4pt}i}$ holds for the mass of emitted inertons.

The value of mass of inertons, which carry out inert and
gravitational properties of particles, has been evaluated in paper
[38]. It has been shown that masses of inertons emitted and then
absorbed by a moving particle are not strongly fixed but
distributed in a wide spectral range much as the photon frequency
varies from zero to the frequency of high-level $\gamma$-photon.

\section{\bf Mass dynamics}

\hspace*{\parindent}
    In our model the canonical particle is considered as a stationary
deformed elementary cell, or superparticle, of the real
space. Hence the origin of the initial particle velocity $\vec
v_0$ is not a moving point in the space as in contemporary
geometry [39] but a cell whose volume is different from that in
the degenerate state. Thereby we can decompose real space in two
subspaces: the external and internal subspaces. Namely, in the
first space the entire cell is considered; it is characterized by
an observable trajectory $l$ and the vector of the particle
velocity $\vec v_0$ which sticks out of the particled cell belongs
to this subspace as well. The second space represents the cell
itself: the cell's size, its inner substructure, degree of its
global surface curvature (or deformation), and hidden motion of
the cell's centre-of-mass. For example, the notion of spin-1/2 \
determined as the proper pulsation of the particle [14] should be
related to the internal space. Thus a moving particled cell is
scattered by surrounding superparticled cells and produces changes
in both the external and internal characteristics.

    So, due to the interaction between a moving particle and
oncoming superparticles the particle undergoes a peculiar
splitting, or fission. The particle mass decays and is spent 
for the creation of inertons. Apparently this is the process that
falls under the study in the internal space.

The particle moves in the external space and its behavior here is
characterized by its proper time $t=l/v_0$, where $l$ and $v_0$ is
the particle trajectory and the particle velocity respectively.
Let us refer all processes which occur in the internal space to
the proper time $t$ of the particle. This means that the dynamics
of a global cellular deformation of the particled cell, i.e., the
dynamics of the inert mass $M$, should be treated as a function of
$t$. Thus two masses may be distinguished in our system: the particle mass
$M$ and the mass of particle's inerton cloud $m$. Besides we have to
operate with the rate of change of mass per time $t$. For this purpose, 
let us introduce values $\dot M$ and $\dot m$, which are the rates 
of change of the particle inert mass and the inerton cloud mass 
respectively.

\subsection{Longitudinal mass}

\hspace*{\parindent}
To describe a dynamics of the particle mass along the particle path $l$, 
we need a model Lagrangian written in the form equivalent to the classical
one. Let us start from the following specific Lagrangian of the internal space,
i.e., the original mass function
\begin{equation}
L_{{\kern 1pt}\parallel}= {\dot M_{\parallel}}^{{\kern 1pt}2}/2  + {\dot m}_{\parallel}^2/ 2 
 -  \frac {\pi} {T} {\kern 3pt}  m_{\parallel}  \dot M_{\parallel}. \label{10}
\end{equation}
Here, the first and the second terms are peculiar kinds of 'kinetic
energies' of the particle inert mass and its inerton cloud mass
respectively; the third term describes the mass exchange; $\pi/T$
is the cyclic frequency of collisions between the particle and the
inerton cloud.

   The Lagrangian (10) of the internal space is similar in its form
to the Lagrangians of the external space used for the particle
motion in papers [12,13,16]. The Euler-Lagrange equations of
motion are as follows
\begin{equation}\label{11}
\ddot M_{\parallel} -\frac \pi T {\kern 1pt}\dot m_{\parallel} = 0,
\end{equation}
\begin{equation}\label{12}
\ddot m_{\parallel} + \frac \pi T \dot M_{\parallel} =0
\end{equation}

If we pay attention to the fact that at the initial moment $t=0$
the total mass was concentrated in the particle, we come to the
solution
\begin{equation}\label{13}
M_{\parallel} = \mu_{\parallel} {\kern 1pt} \cdot| \cos (\pi t/T) |,
\end{equation}
\begin{equation}\label{14}
m_{\parallel} =\mu_{\parallel} {\kern 1pt} \cdot  | \sin (\pi t/T) |
\end{equation}
where the amplitude, i.e. the value of the exchange mass is
\begin{equation}\label{15}
\mu_{\parallel} = M_0 \cdot \big( 1/\sqrt{1-v_0^2/c^2} -1 \big)
\end{equation}
As follows from solutions (13) to (15), the mass of the particle
changes periodically along its path from the initial value 
$M=M_0/\sqrt{1-v_0^2/c^2}$ in node points, 
which corresponds to moments of time $t=nT$ where $n=0, \ 1, \ 2, ...$, 
\ to the rest mass $M_0$ in antinode points.

\subsection{Transversal mass}

\hspace*{\parindent}
To the careful observer, the solutions (13) - (15) correct only along the particle path $l$. However, the particle mass obeys changes also in transversal directions and expression (7) indicates conclusively that its change is different from that prescribed by the solution (13) and (15).

Indeed, if in the longitudinal direction the velocity of inertons is  $\dot x^{\parallel} = v_0$, in transversal directions the velocity is other, $\dot x^{\perp} =c$ (7), and, therefore, in the latter case the Lagrangian must also be distinguished from (10). Namely, it has to look as follows      
\begin{equation}
L_{{\kern 1pt}\perp}= {\dot M}_{{\kern 1pt}\perp}^{{\kern 1pt}2}/2  + {\dot m_{\perp}}^2/ 2 
 -  \frac {\pi} {T} {\kern 3pt}  m_{\perp}  \dot M_\perp \label{16}
\end{equation}
where the first and the second terms describe the 'kinetic
energies' of the particle proper mass and its inerton cloud mass
respectively; the third term describes the mass exchange; $\pi/T$
is the same cyclic frequency of collisions between the particle and the
inerton cloud.

The solutions to the Euler-Lagrange equations derived on the basis of the Lagrangian (16) are     
\begin{equation}\label{17}
M_\perp = \mu_{\perp} {\kern 1pt} \cdot | \cos (\pi t/T) |,
\end{equation}
\begin{equation}\label{18}
m_{\perp} =\mu_{\perp} {\kern 1pt}  \cdot | \sin (\pi t/T) |
\end{equation}
where the amplitude, i.e. the value of the exchange mass is
\begin{equation}\label{19}
\mu_{\perp} = M_0.
\end{equation}

\medskip

Consequently, the particle mass is pumped over from the
particle to the inerton cloud, i.e. to the ambient space, and then
comes back from the cloud to the particle. Thus, the periodical transfer of the particle mass occurs by a tensorial law: it varies from $M\equiv M_0\sqrt{1-v_0^2/c^2}$ to $M_0$ along the particle pass $l$ and changes from $M_0$ to $0$ in transversal directions.

\section{\bf Mass dynamics of inertons}

\hspace*{\parindent}
     In the two previous sections inertons have been treated as
quasi-particles which migrate hoping from superparticle to
superparticle by a relay mechanism. At the same time, it is the
vibrating motion of crystallite's superparticles that knocks
inertons out of the particle. Therefore it is reasonable to
attempt to consider the splitting of the particle mass and its
transformation to the ensemble of inertons from the point of view
of wave process.

The $i$th inerton created by the particle starts to migrate having
a mass $m_{{\kern 0.5pt}i}$. The appearance of the $i$th inerton
means that the superparticle next to the particle takes on the
deformation, on the mass $m_{{\kern 0.5pt}i}$. In other words, the
volume of the corresponding superparticle located at this place is
compressed. So the inerton, i.e. the mass $m_{{\kern 0.4pt}i}$,
starts to barrel in with the initial velocity $\hat c$ from
superparticle to superparticle. Space gradually brakes the moving
inerton and that is why the value of inerton mass $m_{{\kern
0.5pt}i}$ decreases with $r$ passing to zero at distance
$\Lambda_i$ from the particle. At the same time the substrate
elastically wrinkles. In other words, the local deformation is
sensibly transformed into the rugosity [40] of the cellular space
net, which reaches the maximum value at $\Lambda_i$. Then the
space net begins to straighten the rugosity and initiates the
reverse inerton motion: the rugosity grades again into the local
deformation, the inerton takes on the original mass $m_{{\kern
0.5pt}i}$ and after all it arrives to the particle and returns the
mass $m_{{\kern 0.5pt}i}$ to it. It stands to reason that for the
discrete elastic medium the most straightforward pattern for the
motion of the $i$th inerton from the particle's environment can be
presented in Fig. 2.
\begin{figure}
\begin{center}
\includegraphics[scale=2.5]{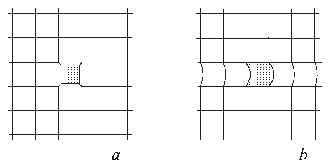}
\caption{ Two limiting cases for the state of an inerton from the
particle's environment in the space net: $(a)$ the deformation,
i.e., the volume change is localized in the cell (here, the
inerton mass $m_{{\kern 0.4pt}i} \ne 0$); \ $(b)$ there is no
deformation in the cell (here, $m_{{\kern 0.5pt}i}=0$, or in other
words, the volume of the cell does not distinguish from that of
nearest cells) while the local deformation is completely
transferred into the rugosity of the space net.} \label{Figure 2}
\end{center}
\end{figure}

In such a way in the system studied we can distinguish two
variables: the inerton mass $m_{{\kern 0.5pt}i}$ that describes a
compressed cell of the space net (Fig 2,$a$) and the space net
rugosity ${\vec u}_{{\kern 0.5pt}i}$ that may be regarded as some
kind of a cell displacement (Fig. 2,$b$). The mass $m_i$ is viewed
as a function of the proper time $t_i$ of the $i$th inerton. The
rugosity is supposed is a function of the radius vector ${\vec
x}_i$ of the $i$th inerton and its proper time $t_{\kern 1pt i}$,
i.e. ${\vec u_{{\kern 1pt }i}}= {\vec u_{{\kern 1pt }i}}({\vec
x}_i, t)$. Further we will deal with dimensionless variables
$\varrho_{{\kern 1pt }i} = m_{{\kern 1pt }i}(\vec x_i, \
t)/m_{{\kern 1pt }i{\kern 0.5pt}0}$ (here $m_{{\kern 1pt }i{\kern
0.5pt}0}$ is the initial maximum value of mass of the $i$th
inerton) and ${\vec \xi}_i={\vec u_{{\kern 1pt }i}}({\vec x}_i, \
t)/(V_0^{\rm sup})^{1/3}$, which may be considered as two
potentials of the inerton.

Let us study a dynamics of the system in the framework of the
following specific Lagrangian
\begin{equation}\label{20}
L = \sum_i \Big[ \frac 12 \ {\dot \varrho}^{{\kern 1pt}2}_i +
\frac 12 \ {{\dot {\vec \xi}}_i}^{{\kern 2.5pt} 2} - {\hat c} \
{\dot \varrho}_{{\kern 1pt }i} {\nabla_i}{\kern 1pt} {\vec \xi}_i
\Big].
\end{equation}
Here, the sum spreads to all emitted inertons; the first two terms
are the rates of change of the mass $\varrho_{{\kern 1pt }i}$ and
the rugosity ${\vec \xi}_{{\kern 0.7pt}i}$ respectively, the third
term describes their interaction. In the Lagrangian (16) the
parameter $\hat c$ specifies the speed of interplay between the
two potentials, namely, the mass potential and the rugosity one.

 The Lagrangian (20) includes the function $\nabla
{\vec \xi}_{{\kern 0.7pt}i}$. In this case, the Euler-Lagrange
equations take the form (see, e.g. ter Haar [41]):
\begin{equation}
\frac {\partial}{\partial {\kern 0.5pt} t} \frac {\partial
L}{\partial {\kern 0.5pt} {\dot Q}} -\frac {\delta L}{\delta  Q}=
0 \label{21}
\end{equation}
where the functional derivative
\begin{equation}
\frac {\delta L}{\delta Q} = \frac {\partial L}{\partial {\kern
0.5pt} Q} - \frac {\partial}{\partial x} \frac {\partial L}
{\partial \bigl( \frac {\partial {\kern 0.5pt}Q }{\partial x}
\bigr) } - \frac {\partial}{\partial y} \frac {\partial L}
{\partial \bigl( \frac {\partial {\kern 0.5pt}Q}{\partial y}
\bigr) } - \frac {\partial}{\partial z} \frac {\partial L}
{\partial \bigl( \frac {\partial {\kern 0.5pt} Q}{\partial z}
\bigr)}. \label{22}
\end{equation}
For the Lagrangian (20) Eqs. (21) and (22) yield the following two
equations:
\begin{equation} \frac{\partial^{{\kern 1pt} 2}\varrho_{{\kern 1pt}i}}
{\partial{\kern 1pt}t^2} - {\hat c} {\kern 1pt} \nabla {\dot {\vec
\xi}_i}=0; \label{23}
\end{equation}
\begin{equation}
\frac{\partial^{{\kern 1 pt} 2}{\vec \xi}{\kern
0.5pt}_i}{\partial{\kern 1pt} t^2} - {\hat c} {\kern 1pt} \nabla
{\dot \varrho}_{{\kern 1pt}i} = 0. \label{24}
\end{equation}
These equations being uncoupled change to the following
\begin{equation}\label{25}
\frac{\partial^{{\kern 1pt}2}\varrho_{{\kern 1pt} i}}{\partial
{\kern 1pt}t^2} -{\hat c}^{{\kern 1pt}2} {\kern 1pt} \nabla^2
\varrho_{{\kern 1pt} i} =0;
\end{equation}
\begin{equation}\label{26}
\frac{\partial^{{\kern 1pt}2}{\vec \xi}_i}{\partial{\kern 1pt}
t^2} - {\hat c}^{{\kern 1pt}2}\nabla\cdot \nabla {\vec \xi}_i = 0
\end{equation}

\noindent  (in general case the right hand sides of Eqs. (25) and
(26) are equal to constants $C_1$ and $C_2$, however, the
appropriate initial and boundary conditions for $\varrho_{\kern
1pt i}$ and $\vec \xi_i$ make it possible to put $C_1, {\kern
1pt}{\kern 1pt}C_2 =0$).

Eq. (25) is a typical wave equation; in our case it describes the
behavior of the mass potential $\varrho_{{\kern 1pt }i}$ of the
$i$th inerton. Thus this characteristic, the mass of the inerton,
indeed changes periodically as we initially suggested. Let us
treat the solution to equation (25), because it is this equation
that is most informative at the study of the mass behavior of the
particle's inertons. The initial conditions are
\begin{eqnarray}
&&\varrho_{{\kern 1pt} i}({\vec x}_{{\kern 0.5pt}i},{\kern 1pt}
0)= \varrho({\vec x}_{{\kern 1pt} i});
\\ \nonumber && \frac {\partial \varrho_{{\kern 1pt} i}({\vec
x}_{{\kern 1pt} i}, {\kern 1pt} 0)}{\partial {{\kern 1pt} t}}= 0.
\label{27}
\end{eqnarray}

\noindent The Cauchy problem to Eq. (25) with conditions (27)
reduces to the problem on oscillations of a string with the fixed
end ${\vec x}_{{\kern 0.5pt}i}=0$. In our case the solution -- in
the form of a standing wave -- will show how the value of inerton
mass decreases along the inerton path, from $ x_{{\kern
0.5pt}i}=0$ to $x_{{\kern 0.5pt}i}=\Lambda_i$.

   Now let us turn to the generalized problem, namely, let us treat
the behavior of the mass of particle's inerton cloud considered as
a single object. The single cloud of inertons is periodically
emitted by the particle and then absorbed again. Besides one
should go on to the proper time $t$ of the inerton cloud and to
the cloud amplitude $\Lambda$ that is found from relationship (2)
(note that such a procedure was already made in papers [13,14] for
other problems). We may describe such a system by the equation
\begin{equation}\label{28}
\varrho_{{{\kern 0.7pt}} tt} - {\hat c}^{{\kern 1pt} 2}{\kern 1pt}
\triangle  {\kern 0.7pt} \varrho =0,
\end{equation}
which is the generalized one to Eq. (25). Here $\varrho$ comprises
the total mass potential of the inerton cloud (and hence the total
inerton mass (11), i.e. $\varrho = m/\mu$). Apparently, the system
studied features the central symmetry in respect to the particle
since inertons issue from the particle throughout the azimuth
angle 2$\pi$ around its path (in a truncated view, it may be
called the radial symmetry of the cloud around the supposedly
motionless particle). Nevertheless the distribution of inerton
trajectories, peculiar rays-strings, and the cause of this
distribution is not discussed herein.

In such a manner we have reduced the problem to the treatment of
wave equation (28) that possesses the central symmetry with the
initial conditions
\begin{eqnarray} \label{29}
&&\varrho{\kern 1pt}(r,\  0) = \varrho{\kern 1pt}(r) = f(r),
\nonumber
\\  && \frac {\partial \varrho{\kern 1pt}(r, \ 0)}{\partial
{{\kern 1pt} t}} = 0; \ \ \ \ \ \ \ \ \
 \end{eqnarray}
the second condition here means that the initial mass of the
inerton cloud is zero (i.e. no any inert mass around a particle in
the beginning). The boundary condition is

\begin{equation}\label{30}
\frac {\partial \varrho}{\partial {\kern 1pt} r} \Big|_{{\kern
1pt} r = \Lambda}=F(\Lambda, {\kern 2pt t}).
\end{equation}

\noindent Because, $\varrho =\varrho {\kern 0.5pt}(r, {\kern 2pt}
t)$, then the Laplase operator in the spherical coordinates with
the center located in the particle changes to the following

\begin{equation}\label{31}
\triangle {\kern 0.7pt}\varrho {\kern 1pt}\rightarrow {\kern
1pt}\frac 1{r}\frac{\partial^{{\kern 1pt}2}} {\partial  {\kern
1pt}r^2}(r\varrho).
\end{equation}

\noindent Therefore equation (28) takes the form of the equation
of radial oscillations

\begin{equation}\label{32}
\frac 1{r}{\kern 0.7pt}(r\varrho)_{rr}=\frac 1{\hat c^{{\kern 1pt
}2}}{\kern 1pt}\varrho_{{\kern 1pt} tt},
\end{equation}

\noindent whose solutions (with conditions (29) and (30)) are
well-known in classical mathematical physics (see, e.g. Refs.
[42,43]). In our case the solution has the form

\begin{equation}\label{33}
  \varrho {\kern 0.7pt} (r, \ t) =   \frac {C}{r}
{\kern 1pt} \Big|\cos\big( \frac {\pi r}{2\Lambda}\big) \Big|
 \Big| \cos\big(\frac {\pi
t}{2T}\big)\Big|
\end{equation}

\noindent and then functions $f(r)$ and $F(\Lambda, {\kern 2pt
t})$ in the conditions (29) and (30) are equal to

\begin{eqnarray}
&&f(r) = C \frac 1{r} \Big| \cos \big( \frac \pi 2 \frac r\Lambda
\big)\Big|,
\\ \nonumber &&F(\Lambda, {\kern 2pt} t) = C \frac{\pi}{2 \Lambda}
{\kern 1pt}\frac{1}{r} \Big| \cos \big( \frac \pi 2 \frac t T
\big)\Big|, \ \ {\rm then} \ \ \varrho_r (r, {\kern 1pt}t)
\big|_{ {r= \Lambda, \atop t=T{\kern 2pt}}} \equiv
F(\Lambda, {\kern 1pt} T) = 0. 
\label{34}
\end{eqnarray}

\noindent The solution (33) can be rewritten explicitly, namely, for two
components of the inerton cloud mass, respective longitudinal $(\parallel{\kern 0.5pt})$ and transversal $(\perp)$:
\begin{equation}\label{35}
 m_{{\kern 1pt}\parallel, {\kern 2pt} \perp}{\kern 2pt}(r, {\kern 2pt} t)= C {\kern 1pt} \frac{\mu_{{\kern 1pt}\parallel,{\kern 2pt} \perp }}{r}
{\kern 1pt} \Big| \cos\big( \frac {\pi r}{2\Lambda}\big) \Big|
  \Big|\cos\big( \frac {\pi t}{2T}\big)\Big|
\end{equation}

\noindent where amplitudes are given in
expressions (15) and (19), i.e., respectively
  $$
\mu_{{\kern 1pt}\parallel} = M-M_0 \equiv M_0 \cdot \big( 1/\sqrt{1-v^2_0/c^2} -1 \big), \qquad  \mu_{{\kern 1pt}\perp} = M_0.
  $$

In equations from (31) to (35) the variable $r$ changes in the range
from the size of the particle, $r=(V_0^{{\kern 0.5pt} \rm
sup})^{1/3}=10^{-30}$ m (or maybe rather $l_{\rm Planck} \sim 10^{-35}$ m), to
the inerton cloud amplitude $r=\Lambda=\lambda {\kern 1pt}\hat
c/v_0$ where $\lambda$ is the de Broglie wavelength.

The solution obtained, (33) to (35), directly demonstrates that
the particle periodically throws about its mass and then
takes it back. The distribution of mass around the
particle follows the amplitude of mass oscillation of the inerton
cloud

\begin{equation}\label{36}
\frac {\mu_{{\kern 1pt}\parallel, {\kern 2pt} \perp}}{r}{\kern 2pt} \cos {\kern 0.5pt}(\pi r/2\Lambda).
\end{equation}

\noindent where $ (V_0^{\rm sup})^{1/3} < r \leq \Lambda$. But
this means that the inerton cloud forms the gravitational
potential of the particle! Indeed, the two orthogonal components  
of mass (36) being distributed around the particle signifies 
the formation of a mass field around the particle. 
In other words,  the mass field is the
result of the defractalization of parts of the particle when its
mass  gets asymmetrically smeared around the "core"
superparticle in the range covered by the inerton cloud amplitude
$\Lambda$ (2).

Thus in this range the cloud's superparticles acquire additional
deformations, i.e. become massive and their mass decreases with
$r$ in compliance with expression (35). Inertons migrating from
the particle and then turning backward to it densely fill the
environment. Superparticles by which inertons migrate contract and
this means that the entire space net around the particle contracts
as well. In  such a manner any test particle having occurred under
the mass field (35) will follow its gradient. That is, it is the
contraction of the space net between two massive particles that
realizes the attraction between them. This is the inner reason of
the phenomenon of gravity, or attraction.

 In a region confined by the size of a superparticle and
the inerton cloud amplitude, $ (V_0^{{\kern 0.5pt}\rm sup})^{1/3}
\ll r \ll \Lambda$, and at $v_{{\kern 0.5pt}0} \ll c$ the
time-averaged mass field (35) is reduced to a good approximation
\begin{equation}\label{37}
m_{\parallel}(r)= (V_0^{{\kern 0.5pt}\rm sup})^{1/3}{\kern 2pt} (v^2_0/c^2) {\kern 2pt}\frac {M_0}{r};   
\end{equation}
\begin{equation}\label{38}
m_{\perp}(r) = (V_0^{{\kern 0.5pt}\rm sup})^{1/3}{\kern 2pt} {\kern 2pt}\frac {M_0}{r}. \qquad\quad
\end{equation}

\noindent Going over to conventional physical units, we have to
multiply the both sides of expressions (37), (38) by a factor
$-G/(V_0^{{\kern 0.5pt} \rm sup})^{1/3}$ where $G$ is
the Newton constant of gravitation; as a results we obtain
\begin{equation}\label{39}
U_{\parallel}(r) = -(v^2_0/c^2) {\kern 2pt} G {\kern 1pt} \frac {M_0}{r}.
\end{equation}
\begin{equation}\label{40}
U_{\perp}(r) = - G {\kern 1pt} \frac {M_0}{r}. \qquad\quad
\end{equation}

Thus the longitudinal component (39) of the gravitational potential of a particle depends on its velocity $v_0$, though the transversal component (40) exactly represents    Newton's gravitational law.

    Consequently, the introduction of inertons means that a new kind of a
mechanics makes its appearance. Actually, in addition to contact
and elastic interactions, which are characteristics of classical
mechanics, and in addition to the specificity of orthodox quantum
mechanics, which repeats the scheme of classical mechanics by using
statistical tools, we gain a field mechanics. This one is based on
the concept of the elastic tessellation space and,
moreover, the concept treats an object as an element of the tessellattice [22].
And just that aspect is responsible for the field mechanics:
inertons  carry not only the momentum and kinetics
energy from one particle to another, but in addition they transfer the local
deformation into the surrounding space. When a test particle is
thrown into the massive field induced by the other particle, it
is contracted and, therefore, the gravitational law (39), (40)
prescribes alterations in energetic and force characteristics of
the test particle caused by the particle contraction due to its
embedding into the contracted tessellation space.

\section{\bf  Discussion}

\hspace*{\parindent}
     In the quantum substrate studied both the substrate itself and its
components -- superparticles-cells are elastic. They are those
peculiarities that enable one to interpret mass as a deformation
of a superparticle. Such a deformation looks like an uniform
reduction of the superparticle's volume (though the deformation is fractal [22]). Peculiarities of the particle motion in the space involve the particle motion itself
including changes in behavior of center-of-mass, i.e. spin
components ($\uparrow,\downarrow$), the motion of the particle's
deformation coat (equivalently a crystallite) along with the
particle, and the creation and migration of inertons. The dynamics
of the system under consideration exhibits the periodical
decay of the particle mass into mass of inertons. At a
distance of $\Lambda$ from the particle the inert mass carried by
the inerton cloud completely turns into the rugosity of the space.
Such periodical transformation of the deformation from the
particle to the spatial rugosity permits the description of the
inerton motion in terms of standing elastic spherical waves. They
have been these waves that induce the deformation, i.e.
gravitational potential $U \propto M/r$ (39), (40) surrounding 
the particle (but the distance $r$ is limited to the enveloping
amplitude $\Lambda$ of the particle's inerton cloud). Therefore,
standing spherical inerton waves are carriers of the real
gravitational interaction between canonical elementary particles.
It is significant that the mutual interaction appears only as a
result of the motion of particles because there is no information
about an absolutely motionless particle relative to the space
beyond the border of the crystallite (see Ref. [14] for details).
In such a manner the crystallite is a peculiar kind of the screen
between the particle and the degenerate space tessellattice.

For instance, the electron's spatial crystallite of the electron 
(the crystallite shields the electron from the degenerate space) 
in the state of rest has the size of the Compton wavelength ${\tilde
\lambda}_{{\kern 1pt}0} =h/M_0c \simeq 2.42 \times 10^{-12}$ m.
The electron of an hydrogen atom is characterized for the lowest
level by the following parameters: $v_{{\kern 0.5pt}0}\simeq 2.1
\times 10^{{\kern 1pt}6}$ m/s, $\lambda =h/M_0v_0 \simeq 0.35$ nm,
and the amplitude of the inerton cloud (2) $\Lambda =
\lambda{\kern 1pt} c/v_0 \approx 20$ nm. Thus the amplitude of the
cloud far exceeds the atom size $a_{{\kern 0.7pt}\rm H} \simeq
0.1$ nm, i.e., $\Lambda/ a_{{\kern 0.7pt} \rm H}\sim 200$. The
crystallite size of the nucleus of a hydrogen atom is determined
by the Compton wavelength of the proton which equals $1.32 \times
10^{-15}$ m. This value is much smaller than all the
above-mentioned parameters. Hence it is reasonable to conclude
that it is not an atomic nucleus that influences the electron, but
on the contrary, the field of transversal potential $U_{\perp}$ (40) of the electron's spherical inerton wave embraces the nucleus. This is evidently
true for the electromagnetic interaction as well, because the
electromagnetic field may be regarded as some kind of a
polarization, which is superimposed on inertons.

    What is a radius of the gravitational potential created by a heavy
material object placed in the degenerate space? It is apparent
that the components of the compound object are in ceaseless
motion. For example, in a solid atoms vibrate in the
neighborhood of their equilibrium positions. These vibrations
result from the atom-atom interaction, which consists of both the
elastic component  of electromagnetic nature and the inertonic one
[18]. Amplitudes of vibrating entities in a solid play a role of
the de Broglie wavelengths of these entities [19]. If, say, a
solid sphere with a radius $R_{{\kern 0.7pt}\rm sph}$ consists of
$N_{\rm sph}$ atoms and the interatomic distance is $\bar a$, the
spectrum of wavelengths of acoustic waves of the sphere is defined
as $2{\kern 1pt}{\bar a}{\kern 1pt} n$ where $ n=1, \ 2, \ ..., \
N_{\rm sph}/2 $. The vibrating motion of atoms generates
respective inerton clouds in the degenerate space, which move in
synchronism with atoms in the acoustic waves. The value of the
corresponding amplitude of the $n$th inerton cloud can be obtained
from the relationship for the frequency of collisions of the
corresponding acoustic wave with the inerton cloud, i.e. amplitude
of inerton cloud $\Lambda_n$ that accompanies the $n$th acoustic
wave is (compare with expression (2))
\begin{equation}
\Lambda_n = 2 {\kern 1pt} \bar a {\kern 1pt} n {\kern 1pt} {c
\over v_{\rm sound}} \label{41}
\end{equation}
where $v_{\rm sound}$ is the sound velocity of the sphere. In the
classical limit the object size far exceeds its de Broglie
wavelength. In that case long wavelength harmonics should
determine the inerton field structure around the object and this
means that any reasonable laboratory time interval $t$ is still
very small in comparison with $T$, so in expression (35) we may
set $|\cos(\pi t/2T)|\cong 1$. For instance, if the sphere has the
volume 1 cm$^3$, \ $\bar a = 0.5$ nm, $N_{\rm sph}= 10^{{\kern
1pt}22}$ atoms, and $v_{\rm sound}= 10^{\kern 1pt 3}$ m/s
expression (41) gives for the biggest amplitude: $\Lambda_{N_{\rm
sph}} \sim 10^{18}$ m. Thus the amplitude of the longest inerton
wave $\Lambda_{N_{\rm sph}}$, which the material sphere generates
as a whole is the maximum distance to which the inerton field of
the sphere propagates in the form of the standing spherical
inerton wave. 

Thereby, we can write the gravitational potential of
a compound spherical object, which is similar to the particle's (40),
\begin{equation}\label{42}
U=-GM/r   
\end{equation}
where $r$ is limited by inequalities $R_{\rm sph}\leq  r \ll
\Lambda_{N_{\rm sph}}$. Let us roughly estimate the time necessary
for the emitted inertons to return. For simplicity we assumed that
the velocity of inertons $\hat c =c$. Then expressions above yield
for the corresponding time, namely, for time wave period:
$T_{N_{\rm sph}}= \Lambda_{N_{\rm sph}}/c \sim 300$ years! The longitudinal gravitational component proportional to $v_0^2/c^2$ is also available around macroscopic objects, though its manifestation will be treated in a separate work.   

   On the other hand, the description of the quasi-stationary
potential (34$a$) around the object can mathematically be presented
in the form of metric tensor components $g_{ij}$, which are
fundamental in Einstein's general relativity, because of the
alteration of the size of superparticles surrounding the central
object. Such size alteration we associate with the induction of
the gradient of the deformation field (i.e. inerton filed).
Consequently at the macroscopic approximation the metric tensor of
the deformation potential of the degenerate space may really be
regarded as an effective gravitational one.

    We also enlarge on special relativity. The microscopic mechanism
forming the basis for Lorentz transformations in the space
substrate has been analyzed by the author in paper [13]. In
essence, the microscopic theory takes into account that any
material object consists of elementary particles and  each
particle is surrounded by its own deformation coat, equivalently,
the crystallite. The behavior of the crystallite has been
described in terms of the discrete hydrodynamics where the point
is regarded to be equal to the size of the crystallite, i.e., the
Compton wavelength ${\tilde \lambda}_{{\kern 1pt}v_{_0}}$. With
the whole object, one may introduce some effective crystallite as
well. Then the size of the object crystallite will be defined by
the object's Compton wavelength $\lambda_{{\kern 1pt}\rm Comp}$.
However in the limit $l_{{\kern 0.7pt}\rm obj} \gg \lambda_{\rm
Comp}$ (for space) where $l_{\rm {\kern 1pt }obj}$ is the object
typical size and $t\gg \lambda_{{\kern 1pt}\rm Comp}/v_0$ (for
time) where $v_0$ is the object velocity, the discrete
hydrodynamics can be easily replaced by the kinematics of special
relativity which correctly depicts all the parameters of the
object but does not reflect the actual (microscopic) origin that
is covert by the relativity formalism.

\section {Concluding remarks}

\hspace*{\parindent}
        Several important aspects of motion of a canonical
particle have been studied in the present paper. The study based
on the submicroscopic approach has allowed  the consideration of
behavior of a particle and the surrounding space along the
whole particle path. It has been shown that the availability of
the deformation coat, i.e. crystallite, whose massive
superparticles are found in the ceaseless vibrating motion plays
the role of an original generator that knocks inertons out of the
particle, as a result of which its mass decays 
from $M_0/\sqrt{1-v_0^2/c^{\kern 1pt 2}}$ to $M_0$ along the particle path 
and from $M_0$ to zero in transversal directions. Then the elastic space 
gives back inertons to the particle restoring its total mass. All
these changes occur within  the spatial period of particle
oscillation, i.e., the de Broglie wavelength $\lambda$. Thereby we
can say such mechanism represents a real {\it perpetual motion machine}, which 
is launched by an initial push transmitting the velocity ${\vec v}_0$ to the particle.

     It is apparent the particle's inerton migrating in the space is specified by coordinates and the velocity and the equation of motion
\begin{equation}\label{43}
\ddot r + \frac {\gamma}{m}{\kern 1pt}r = 0
\end{equation}
gives the information about these parameters [12-14]. Yet the
inerton as a quasi-particle carries the local deformation
$\varrho$ whose migration in the degenerate space is subjected to
wave equation (25)
\begin{equation}\label{44}
\varrho_{{\kern 1pt}tt} - {\hat c}^{{\kern 1pt}2} \triangle {\kern
0.5pt} \varrho = 0.
\end{equation}
A couple of equations (43) and (44) completely describes the
behavior of the inerton. The first equation, (43), describes the
motion of the "core" of the inerton; this equation enables the
submicroscopic interpretation of quantum mechanics. The second
one, (44), describes the value of deformation that the inerton
transfers from the particle into the tessellattice: the value
$\varrho$ decreases by the law of inverse distance $1/r$ where $r$
is the space between the particle and the inerton (note that $r$
obtained from Eq. (43) is entered into Eq. (44)). Practically
speaking, this allows the interpretation of phenomenon of
attraction as a contraction of the space tessellattice between material
objects.

This can readily be visualized by the following pattern. A
physical point being embedded into a medium radiates standing
acoustic (longitudinal  type) radial waves which spread down to
the distant spherical surface; hence, the medium vibrates in the
range covered by the sphere's radius. A standing wave, as is
well-known, is characterized by ranges of stretching and
contraction of the medium. Thus our central point should be
treated as a node point and the distant spherical surface's points
to be antinode points, that is, the medium is contracted in the
vicinity of the central point and stretched at the distant surface
(see Fig. 2,b). As a result, the gradient of tension will exactly
be directed to the central point simulating the phenomenon of
classical "static" gravity.

Now let $r$ be a space between two particles and let it obeys the
inequality $r<(\Lambda_1+ \Lambda_2)$ where $\Lambda_{1(2)}$ are
envelope amplitudes of inerton clouds of the two particles. In
this situation the particles will fall under the behavior
predicted by Newton/Coulomb law. This has been shown in the
previous Section. However, such kind of the interaction between
canonical particles is correct only when velocities of
interacting particles are widely different.

     If velocities of particles have the same order,
the pattern of particle-particle interaction radically changes. In
this case the elasticity $\gamma$ of each of the inerton clouds is
approximately identical in value and hence the particles will
contact each other through the attractive inerton potential that
obeys the harmonic law, $\frac 12 {\kern 1pt} \gamma {\kern 0.5pt}
r^2$. Thus if elasticity constants $\gamma_1$ and $\gamma_2$ of
the two inerton clouds have the same order, one can regard that
the interaction between the particles is harmonic. (Indeed, let in
a many particle system one particle moves towards the other one
and let absolute value of their velocities be the same. Denote the
mean mass and the mean velocity of particles' inertons as $\langle
m \rangle$ and $\langle v \rangle$ respectively. Then at the
distance between the particles lesser $(\Lambda_1 +\Lambda_2)$
inertons from one cloud begin to contact inertons from the other
one. This means that inertons coming from the opposite directions
should elastically collide as the absolute value of their momenta
is the same, $\langle m \rangle \langle v \rangle$. Therefore the
two particles owing to their inerton clouds will contact each
other much like two elastic balls. The particles repulsing move in
opposite directions and again interact with similar particles
which return these two to their initial positions).

The existence of the harmonic inerton potential in an ensemble of
particles has experimentally been substantiated for atoms in the
crystal lattice of a metal [18] and quite recently for the
KIO$_3\cdot$HIO$_3$ crystal, in which just the harmonic inerton
potential $\frac 12 {\kern 1pt}\gamma {\kern 0.5pt} r^2$ of
hydrogen atoms provides for their clustering [19]. Besides it has
theoretically been shown in paper [44] that it is the inerton
potential $\frac 12 {\kern 1pt}\gamma {\kern 0.5pt} r^2$ of
nucleons that holds them in a nucleus; in other words, the inerton
field of nucleons is a real confining field that ensures the
nuclei stability.

    Similarly, one can set the criterion regarding the interaction
of classical objects. Let there be two the same solid spheres,
which are characterized by a temperature $\Theta$. This
temperature is directly defined by the mean kinetic energy of the
spheres' atoms. From the relationship $\frac 12 {\kern
1pt}Mv^2_{\Theta}= \frac 32 {\kern 0.7pt}k_{\rm B} \Theta$ one
gets the thermal velocity of vibrating atoms $v_{\Theta}=
\sqrt{3{\kern 0.5pt}k_{\rm B}\Theta /M}$. Then one can gain the
corresponding de Broglie mean thermal wavelength $\lambda_{{\kern
1pt}\Theta}=h/Mv_{\Theta}=h/\sqrt{3{\kern 0.5pt}k_{\rm B}\Theta
M}$ and further, in agreement with relationship (2), obtain the
mean thermal amplitude of the inerton cloud $\Lambda_{\Theta}=h
c/(3{\kern 0.5pt}k_{\rm B}\Theta)$ of the atom. If the distance
$r$ between surfaces of the spheres meets the inequality
$r<2\Lambda_{\Theta}$, a noticeable  elastic inerton correlation
will be induced between the spheres, i.e. the spheres will suffer
the inerton attraction with the potential $\frac 12 {\kern 1pt}
\gamma {\kern 0.5pt} r^2$. For example, $\Lambda_{\Theta} \sim 10$
$\mu$m at the room temperature and $\Lambda_{\Theta} \sim 1$ mm at
4 K.  As was mentioned above, such inerton correlations in a solid
were studied in paper [18].

It is important to note that in the experimental study of
short-range gravitational effects one should leave the induced
electromagnetic influence out of the gravitational action. In
other words, the van der Waals interaction between macroscopic
objects (see, e.g. reviews [45-47]) should be reduced. (The effect
of Casimir force at a distance about several $\mu$m has recently
been studied theoretically by Lambrecht et al. [48,49]. Then Long
et al. [50] based on the method developed in Refs. [48,49] have
calculated the Casimir background and concluded that a
gravitational-strength Yukawa force should be distinguishable from
the Casimir one at the scale of about 3 $\mu$m; however, any new
gravitational force has not been revealed experimentally in the
range between 75~$\mu$m and 1~mm.)

      At $r>2\Lambda_{\Theta}$, thermal inerton correlations
between objects are absent and this is why one can simulate the
interaction based on the notion of the whole inerton cloud that
oscillates in the space around each of the objects. In this case,
as it has been shown above, the Newton law is realized. A
deviation from the Newton law uncovered by general relativity in a
macroscopic range needs a separate analysis of the paired
interaction of two attracting objects. Nevertheless, the problem
does not occur challengeable from the point of view of the
developing submicroscopic concept.

    The manifestation of the inerton field is observed not only in
gravitation induced by this field. Inerton waves are directly
responsible for the inexplicable force that since the old times
has been called the force of inertia. Many times each of us has
undergone the influence of this force, which continue to move
everybody when he/she tries to stop or turn abruptly. In such
cases our own clouds of inertons continue pushing us slightly when
we stop after a fast movement. The inerton field solves also the
problem of centrifugal force, which so far still remained an
understandable phenomenon of classical mechanics! It is obvious
from the stated above that the centrifugal force, which acts on a
body moving along a curve line, should appear as a response of the
space on a centripetal acceleration applied to the body.

\vspace{6mm} \hspace{1cm}  {\bf Acknowledgement}

\vspace{3mm} I am deeply indebted to late Professor Michel Bounias
for the critical reading of the manuscript and the valuable
remarks.


\begin{thebibliography}{99}
\bibitem{40} The rugosity of the space net implies some kind of
            a deformation as well, but this is the deformation of
            the entire cellular ordering. To avoid confuse, we will
            use the term "rugosity" in an attempt to distinguish
            it from the local deformation of the net, i.e.
            the deformation of superparticle. Perhaps in pure physics
            terms this notion one can indentify with the tension, or
            stress, of the lattice.
\end{thebibliography}
\end{document}